# A Half-analytical Elastic Solution for 2D Analysis of Cracked Pavements

H. Nasser and A. Chabot

Institut Français des Sciences et Technologies des Transports de l'Aménagement et des Réseaux, Materials and Structures Department, Bouguenais, France.

**Abstract**

This paper presents a half-analytical elastic solution convenient for parametric studies of 2D cracked pavements. The pavement structure is reduced to three elastic and homogeneous equivalent layers resting on a soil. In a similar way than the Pasternak's modelling for concrete pavements, the soil is modelled by one layer, named shear layer, connected to Winkler's springs in order to ensure the transfer of shear stresses between the pavement structure and the springs. The whole four-layer system is modelled using a specific model developed for the analysis of delamination in composite materials. It reduces the problem by one dimension and gives access to regular interface stresses between layers at the edge of vertical cracks allowing the initial debonding analysis. In 2D plane strain conditions, a system of twelve-second order differential equations is written analytically. This system is solved numerically by the finite difference method (Newmark) computed in the free Scilab software. The calculus tool allows analysis of the impact of material characteristics changing, loads and locations of cracks in pavements on the distribution of mechanical fields. The approach with fracture mechanic concepts is well suited for practical use and for some subsequent numerical developments in 3D.

**Keywords:** Calculus tool, half-analytical solution, M4-5nW, pavement, cracks, debonding.

# 1 Introduction

Most of pavement structures subjected to the climatic hazards and repeated passage of heavy loads (bus, truck, aircraft, etc.), reach their limit of life and need rehabilitation. They can be reinforced by several methods such as adding or replacing layers. Depending on the type of roads, these layers are made with different types of materials such as recycling bituminous material, cement concrete layers especially in France for urban area and/or glass grids. Taking into account the possibility of presence of discontinuities in a pavement structure is then important for analysing its final service life and predicting its failure mode [1]. Nowadays, these reinforcements are poorly mastered because they still are generally proposed empirically or on the basis of design methods that do not take into account partial discontinuities at the interfaces between layers and/or vertical cracks in multilayer pavement structures. Indeed, most of the design pavement methods, such as the French method [2], use the elastic Burmister's axisymmetric model [3]. This modelling as well as advance ones developed for 3D viscoelastic pavement tools, for instance ViscoRoute© [4-6], cannot take into account vertical discontinuities and/or partial delamination encountered in these multilayer damaged structures. So far, there is no tool for engineers to calculate and analyse the mechanical fields responsible of the debonding between layers in the cracked pavement structures. The main objective is to propose practical tools for the design of appropriate and sustainable solutions of pavement reinforcement. To be widely used in engineering offices, these tools must be fast and easy to implement in software.

The analysis of multilayer structures partially cracked may be difficult due to singularities located at the interfaces between layers of material near the edges or vertical cracks [7-8]. Many works, preliminary on theoretical modelling [9-11], then on experimental developments [12-13] and on numerical modelling [14] may help to understand how taking into account those discontinuities in such 3D multilayer structures. Among the existing numerical methods, the Finite Element Method (FEM) is nowadays the most used one. This method has many advantages such as introducing complex boundary conditions. But it requires the use of fine meshes near cracks or discontinuities. It increases the work of 3D meshing and the time of calculation. At the end of 90s, advanced numerical models such as G-FEM [15] and X-FEM [16] have been developed in order to overcome the problems of re-meshing during crack propagation. More recently, new models such as TLS (Thick Level Set) approach, developed by [17], permit the description of initiation and propagation of defects in a unified framework [18]. However, the introduction of partial inter-facial cracks is not fully established yet, even if there is many promising works on this subject [19-20]. In addition, all these models and approaches to solve the problem may be too heavy to be contained in practical software.

In an alternative way, the French Institute of Science and Technology of Transport, Development and Networks (IFSTTAR) proposes an approach that is used in this paper. It uses one of the multi-particle models of multilayer materials (M4) that have been specially developed to study the edge effects and delamination in composite structures materials [21-23]. In such models, assumptions for the unknown field variables are introduced for each layer separately.

The M4 selected herein for the pavement bending problem contains five kinematic fields (5n) per layer i (i ∈ {1, n}, where n denotes the total number of layers). The M4-5n is formulated by making approximations on the stress field. It has a linear polynomial expansion through the thickness, in each layer, for the bending stresses. The Hellinger–Reissner variational principle [24] is used. Then from a multi-particulate description in 2D, this modelling approach can determine the intensity of the 3D mechanical fields. As opposed to other classical models, these mechanical models yield finite stresses at a free edge or crack tip at the interface point location of two different layers. While the number of unknowns may be significant, the semi-analytical solution of the equations of the model allows for easy and quick parametric studies. It is useful for modelling office engineer calculation types [25-26].

For pavement-cracked applications, an initial adaptation of the multi-particular model (M4-5n) combined with the Boussinesq model (B) [27] for the soil has been proposed in [28-29]. The M4-5nB proposed has shown its effectiveness in modelling 3D mechanical fields in the case of loading with or without introduction of thermal gradients. However, the numerical solution of the system of differential equations of order 2 of the M4-5n equations coupled to the integral Boussinesq's equations still takes too long for the expected final 3D tool. A faster solution using Winkler springs [30] for the soil, named M4-5nW, was tested recently, and applied to the case of plane strain of bilayer and tri-layer pavement [31]. The results showed that the stress fields at the interfaces are similar to those obtained by the simulation of a structure with a Boussinesq soil and the FEM solution far from the interface between the pavement and the soil. In that case the M4-5nW solution is obtained four times CPU faster than the M4-5nB and five times faster than the one obtained by the FEM.

Following these previous works [29] and in order to simplify to a maximum the modelling of the real 3D pavement, this one is finally chosen equivalent to 3 layers (surface course, base course and sub-base course) [32] resting on a soil. In the aim to get better approximations of the mechanical fields between the pavement layers and the soil, the modelling of the soil is improved in this work. Similar to Pasternak's assumptions [33], the soil is taken equivalent to a combination of a fictitious layer (shear layer) ensuring the transfer of shear stresses between the sub-base course and Winkler's springs. The all four layers are then modelled by means of the M4-5n.

This paper is based upon Nasser's work presented in [34]. In addition it includes the full and detailed method to construct the "2D plane strain reference case" and the calculus of the elastic stress energy of the M4-5n. In the first section, the specific elastic model denoted finally M4-5nW is presented. In the second section, a composite pavement structure is studied in the aim to determine and to optimize the thickness of the equivalent layer that is added to the Winkler's massif soil. The third section illustrates the advantages of such modelling for a 2D cracked pavement case. Few parametric calculations to determine the most critical load position relative to the existence of a vertical crack and to examine some thermal changing effects in a composite pavement structure are finally illustrated on the distribution of some interface stresses.

## 2 Development of M4-5nW for the pavement structures

In this section, the Multi-Particle Model of Multilayer Materials (M4-5n) with 5n equilibrium equations (n: total number of layers) is presented. Then the modelling for the soil is discussed in the aim to determine the thickness of the shear layer.

## 2.1 The M4-5n

### 2.1.1 General system of the M4-5n equations

The M4-5n is adopted to simulate bending problems in pavement structures [29]. It belongs to the M4 family [21-23]. Assuming that kinematic fields and stress fields may be written per layer rather than per the total thickness of the multi-layered, the M4 family estimates the mechanical fields are thinner than the "Layer Wise plate models" [35]. The M4-5n construction is based on a polynomial approximation by layer in z for the in-plane stress fields (x and y represent the coordinates of the plane of the layers and z represents the vertical coordinate). The thickness of each layer is given by $e^i = h^i_+ - h^i_-$, where $h^i_+$ and $h^i_-$ are the ordinate of the higher and lower face of layer i respectively (i ∈ {1,n} where n is the total number of layers). The coefficients of these polynomial approximations are expressed via Reissner's classical stress generalized fields. The shear and normal stresses (respectively $\tau^{i,i+1}_\alpha$ (x,y) and $v^{i,i+1}(x,y)$, α ∈ {1,2}, i ∈ {1,n-1}) at the interface between layers i and i+1 (similarly i-1 and i) are ensuring the continuity of the 3D stress field, $\sigma_{kl}$ (k ∈ {1,3}, l ∈ {1,3}), between these two consecutive layers (Eq. 1-2).

$$\tau^{i,i+1}_\alpha(x,y) = \sigma_{\alpha 3}\left(x,y,h^i_+\right) = \sigma_{\alpha 3}\left(x,y,h^i_-\right) \quad (1)$$
$$v^{i,i+1}(x,y) = \sigma_{33}\left(x,y,h^i_+\right) = \sigma_{33}\left(x,y,h^i_-\right) \quad (2)$$

Where $\tau^{0,1}_\alpha(x,y), v^{0,1}(x,y)$, $\tau^{n,n+1}_\alpha(x,y)$ and $v^{n,n+1}(x,y)$ represent respectively the boundary conditions above and below the pavement interface related to interface efforts between the multilayer structure and its external environment.

This model can be viewed as a superimposition of n Reissner plates linked by interfacial forces. Between two adjacent material layers, it becomes possible to express delamination criteria in terms of interfacial forces [26, 36]. The evaluation of interface stresses is obtained by a method based on the Hellinger–Reissner variational principle [24]. This formulation reduces then the real problem 3D at the determination of plane fields (x, y) per each layer i and interface i, i+1, (and i-1, i). These fields in the plane (x,y) are regular. Thus the real object 3D (2D) is transforming into one geometry 2D (1D). Furthermore, the M4 approach avoids singularities by giving a finite value of stresses at plate edges [23].

In order to simplify the analysis, the equations of M4-5n used to model the three equivalent pavement layers and the shear layer ensuring the connection between the pavement and the Winkler's soil (Figure 1), are solved here under the assumption of plane strain conditions and assuming that the volume forces are negligible. Subsequently, the mechanical fields of the M4-5n depend only on variable x.

The layers are numbered from top to bottom of the structure. We denote $E^i$, $e^i$ and $v^i$ (i ∈ {1, 4}) respectively the Young's modulus, the thickness, and the Poisson's ratio of each layer i. $E^s$ is the Young's modulus of the soil, and k is the stiffness of the springs. The uniform pressure load is assumed to be applied vertically on the pavement by help to the normal stress $v^{0,1}(x)$; the shear stress at the interface between the first layer of the pavement and its external environment is null ($\tau_1^{0,1}(x) = 0$).

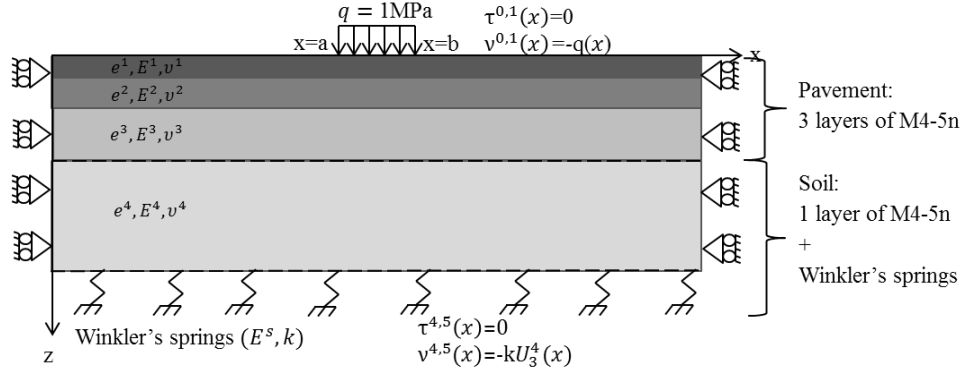

Figure 1: The equivalent pavement structure modelled by the M4-5nW

After combining the mechanical M4-5n equilibrium and behaviour equations per layer i, a system of three 2$^{nd}$ order differential equations for each layer i (i ∈ {1, n}) of the pavement multilayer is generated (Eq. 3-5).

$$\frac{e^i E^i}{1-v^{i^2}} U_{1,11}^i(x) = \tau_1^{i-1,i}(x) - \tau_1^{i,i+1}(x) \tag{3}$$

$$\frac{e^{i^2} E^i}{5(1-v^{i^2})} \phi_{1,11}^i(x) - U_{3,1}^i(x) - \phi_1^i(x) = \frac{1+v^i}{E^i}(\tau_1^{i-1,i}(x) + \tau_1^{i,i+1}(x)) \tag{4}$$

$$U_{3,11}^i(x) + \phi_{1,1}^i(x) = \frac{12(1+v^i)}{5e^i E^i}(v^{i-1,i}(x) - v^{i,i+1}(x)) - \frac{(1+v^i)}{5E^i}(\tau_{1,1}^{i-1,i}(x) + \tau_{1,1}^{i,i+1}(x)) \tag{5}$$

$U_1^i(x)$, $U_3^i(x)$ and $\phi_1^i(x)$ are respectively the average in-plane displacements, average out-plane displacements and average rotations of the layer i. Similarly, a system of two 1$^{st}$ order differential equations per interface i,i+1 (i ∈ {1, n-1}) is obtained as follows (Eq. 6-7):

$$\frac{e^{i+1}}{12} U_{3,1}^{i+1}(x) + \frac{e^i}{12} U_{3,1}^i(x) + U_1^{i+1}(x) - U_1^i(x) - \frac{5e^{i+1}}{12}\phi_1^{i+1}(x) - \frac{5e^i}{12}\phi_1^i(x) =$$
$$-\frac{e^{i+1}(1+v^{i+1})}{12E^{i+1}}\tau_1^{i+1,i+2}(x) - \frac{e^i(1+v^i)}{12E^i}\tau_1^{i-1,i}(x) + \left(\frac{e^i(1+v^i)}{4E^i} + \frac{e^{i+1}(1+v^{i+1})}{4E^{i+1}}\right)\tau_1^{i,i+1}(x) \tag{6}$$

$$U_3^{i+1} - U_3^i(x) = \frac{9e^{i+1}}{70E^{i+1}}v^{i+1,i+2}(x) + \frac{9e^i}{70E^i}v^{i-1,i}(x) + \frac{13}{35}\left(\frac{e^i}{E^i} + \frac{e^{i+1}}{E^{i+1}}\right)v^{i,i+1}(x) \tag{7}$$

The manipulation of the three 2$^{nd}$ order differential equations per layer i and two 1$^{st}$ order differential equations per interface i,i+1 leads to a system of 3n 2$^{nd}$ differential equations of M4-5n. This system function of (x) is written as given by (Eq. 8):

$$AX''(x) + BX'(x) + CX(x) = DY^{0,1'}(x) + EY^{n,n+1'}(x) + FY^{0,1}(x) + GY^{n,n+1}(x) \tag{8}$$

We denote respectively « ' » and « '' » the first and the second derivatives of the unknown fields with respect to the variable x. X(x) is the vector containing the Reissner-Mindlin's average kinematic unknowns by layer. $Y^{0,1}(x)$ and $Y^{n,n+1}(x)$ are the interface efforts vectors between the multilayer structure and its external environment (Eq. 9). These vectors are used to write the boundary conditions of vehicle's load exerted on the top and below of the pavement structure. The indexes " 0,1 " and " n,n+1 "are used respectively for the boundary locations between outside and the first layer and between the outside and the fourth layer where it provides the connection between the shear layer and the Winkler's massive.

$$X = \begin{bmatrix} X^1 \\ \vdots \\ X^n \end{bmatrix}_{3n\times 1} \;;\; [X] = \begin{bmatrix} U_1^i \\ \phi_1^i \\ U_3^i \end{bmatrix} \;;\; Y^{0,1} = \begin{bmatrix} \tau_1^{0,1} \\ \nu^{0,1} \end{bmatrix}_{2\times 1} \;;\; Y^{n,n+1} = \begin{bmatrix} \tau_1^{n,n+1} \\ \nu^{n,n+1} \end{bmatrix}_{2\times 1} \quad (9)$$

The matrixes $[A]_{3n\times 3n}$, $[B]_{3n\times 3n}$, $[C]_{3n\times 3n}$, $[D]_{3n\times 2}$, $[E]_{3n\times 2}$, $[F]_{3n\times 2}$ et $[G]_{3n\times 2}$ of the system (8) depend only on geometric and mechanical parameters of equivalent elastic problem. The use of the Mathematica software allows writing analytically these matrixes in order to reduce the number of operations to perform. It minimizes thus the computing time during the numerical resolution of the problem. The filling shape of these matrixes is given in Appendix A.

The boundary condition systems of the multilayer edges and cracks are expressed as function of both the kinematic unknowns and interface forces using the constitutive equations of layer i for M4-5n [21]. Introducing cracks that are either longitudinal (x direction) or vertical over the thickness of one or more layers (z direction), requires considering that crack tips constitute two free edges, whose distance interval from one another corresponds to the width of the crack. The boundary conditions are expressed in the form of the five equations per layer system. The multilayer pavement is considered blocked at its edges, far from the loading where the material is confined (Eq. 10).

$$\begin{cases} \lim_{x\to\pm\infty} U_{1,11}^i = 0 \\ \lim_{x\to\pm\infty} \phi_1^i = 0 \\ \lim_{x\to\pm\infty} (\phi_1^i(x) + U_{3,1}^i(x) + \frac{1+\upsilon^i}{5E^i}(\tau_1^{i-1,i}(x) + \tau_1^{i,i+1}(x))) = 0 \end{cases} \quad (10)$$

The boundary conditions of free edges (Eq. 11) are applied to represent vertical cracks.

$$\begin{cases} \lim_{x \to \pm\infty} U^i_{1,1} = 0 \\ \lim_{x \to \pm\infty} \phi^i_{1,1} = 0 \\ \lim_{x \to \pm\infty} (\phi^i_1(x) + U^i_{3,1}(x) + \frac{1+\upsilon^i}{5E^i}(\tau^{i-1,i}_1(x) + \tau^{i,i+1}_1(x))) = 0 \end{cases} \quad (11)$$

The comparison between the results of the M4-5n with those from other types of modelling can be done on the value the elastic energy $W^{5n}_{2D}(x)$. After many simplifications, it is given below analytically in terms of the material characteristics and the unknown kinematics fields of layers i and interface stress field (Eq. 12).

$$W^{5n}_{2D}(x) = \sum_{i=1}^{n} \begin{bmatrix} \dfrac{e^{i^5} E^i (1+\upsilon^i)}{120(1-\upsilon^{i^2})^2} \int_0^L \phi^{i^2}_{1,11}(x)dx + \dfrac{7e^{i^3}}{60(1-\upsilon^i)} \int_0^L \phi^i_{1,11}(x)(\tau^{i-1,i}_1(x) - \tau^{i,i+1}_1(x))dx \\ + \dfrac{e^i E^i}{2(1-\upsilon^{i^2})} \int_0^L U^{i^2}_{1,1}(x)dx + \dfrac{e^{i^3} E^i}{24(1-\upsilon^{i^2})} \int_0^L \phi^{i^2}_{1,1}(x)dx \\ + \dfrac{9e^i(1+\upsilon^i)}{15E^i} \int_0^L (\tau^{i-1,i}_1(x) - \tau^{i,i+1}_1(x))^2 dx \\ + \dfrac{17e^i}{280E^i} \int_0^L (v^{i,i+1}(x) - v^{i-1,i}(x))^2 dx + \dfrac{e^i}{8E^i} \int_0^L (v^{i,i+1}(x) + v^{i-1,i}(x))^2 dx \end{bmatrix} \quad (12)$$

**2.1.2 Numerical solution using the Finite Difference Method**

The numerical solution of the system (Eq. 8) is done by applying a non-dimensional method along with the finite difference method in order to avoid any problems of ill conditioning of matrixes during numerical manipulations. Then, the studied multilayer equivalent medium is discretized to N points (nodes), in x direction, according to the Newmark scheme [37] which is implemented in the French open source software for numerical computations known as Scilab (http://www.scilab.org/fr). Newmark's method allows an easily introduction of a vertical crack across a layer because each layer of M4-5n has its own equilibrium equations (Eq. 3-5), constitutive law and boundary conditions (Eq. 10-11). The introduction of such type of cracks is done by replacing, in the final systems, corresponding lines and columns between two consecutive nodes by the boundary conditions of the free edges. The filling shape of the matrixes corresponding to a case with a vertical crack in the third layer is given in Appendix B. After several manipulations, it is possible to reduce the order of the system of differential equations (Eq. 8) to obtain the final system (Eq. 13).

$$\mathbb{A}\mathbb{X}(x) = \mathbb{B}\mathbb{Y}^{0,1}(x) + \mathbb{C}\mathbb{Y}^{n,n+1}(x) \quad (13)$$

Where $[\mathbb{A}]_{3nN \times 3nN}$ is the resultant matrix, $[\mathbb{B}]_{3nN \times 2N}$ and $[\mathbb{C}]_{3nN \times 2N}$ represent the tensor forces and other boundary conditions exerted above and below the multilayer via vector $\mathbb{Y}^{0,1}(x)$ and $\mathbb{Y}^{n,n+1}(x)$ respectively. Whatever the case of a load position with respect to the crack, the multilayer thus modelled by the M4-5n is divided into four wired zones (Figure 2).

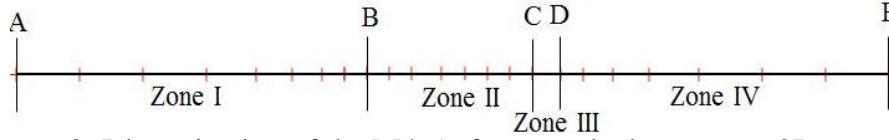

Figure 2: Discretization of the M4-5n for a cracked pavement 2D structure

- The zone I, containing the NI nodes, is located to the left of the load with a decreasing preference mesh from the left toward the right side;
- The zone II, containing the NII nodes, contains the loading zone, its discretization is regular;
- The zone III, containing 2 nodes only, corresponds to the cracked zone (where the M4-5n boundary conditions of free edges are written between two consecutive nodes);
- The zone IV, containing the NIV nodes, is located to the right of the load with an increasing preference discretization from the left toward the right side

To determine the number of nodes for each zone, a test of convergence of the mechanical fields is done.

## 2.2 Winkler's solution for the soil mass

Although the modelling of the soil by a massive of springs (Winkler's massive) [30] is simple and fast to use, it cannot takes into account the shear stresses between the soil and the pavement structure. It considers that they are equal to zero when in reality they are not [31]. To improve the modelling and get better approximations of the mechanical fields near the soil, and in the aim to keep the simplicity of this modelling, a M4-5n layer is added between the pavement structure and the Winkler's springs in order to ensure there the transfer of shear stresses (see Figure 1). The stiffness of springs, k, is then extrapolated to four layers according to the Odemark formula as given by (Eq. 14) [38], where h*, the equivalent thickness of the multilayer, is calculated from the "Method of Equivalent Thickness " (MET). f is a correction factor which is equal to 0.9 for a bilayer. For an upper number of layers, f is equal to 0.8 except for it first interface. For the first interface f is equal to 1.0 or 1.1 if the radius of the load is greater than the thickness of the layer i.

$$k = \frac{E^s}{h^*} \; ; \; h^* = \sum_{i=1}^{i=n} f\, e^i \sqrt[3]{\frac{E^i}{E^s}} \tag{14}$$

Marchand and al. [39] found that the formula (Eq. 14) can be used only if the thickness of the layers is greater than half the radius of the load or if the ratio $\frac{E^i}{E^{i+1}}$ is greater than 2. The assumption of continuity of vertical displacements between the multilayer (n layers) and the Winkler's soil (indexed by n+1) implies the boundary conditions the interface M4-5n stress fields between the fourth layer and Winkler's springs (Eq. 15).

$$\begin{cases} \tau_1^{n,n+1}(x) = 0 \\ v^{n,n+1}(x) = -kU_3^n(x) \end{cases} \quad (15)$$

In the development of the simplified modelling tool M4-5nW the thickness of the "shear layer" must be determined. This thickness must respect two rules:
- The geometric validity of the assumptions of plate models ($e^i \ll L$). A ratio of 10 between the length of the pavement (L) and the thickness of the shear layer ($e^n$) is considered sufficient to keep the assumption of plate models ($e^n \leq L/10$).
- Due to the assumptions made in the M4 constructions, a thickness relationship between two consecutive layers of the M4-5n ($e^n/e^{n-1} \leq 4$) must also be taken into account in order to obtain results sufficiently accurate [28].

## 2.3 The resulting M4-5nW approach

The resulting model is a pavement structure reduced to three elastic and homogeneous equivalent layers resting on a soil modeled by adding a shear layer to Winkler's springs (Figure 1). This pavement structure is modeled using the M4-5n. In the modelling, each pavement layer has its own boundary conditions and the introduction of a macro-vertical crack in one layer only is easy to be introduced separately. In the aim to consider real size of pavement structures and to offer interesting computational power, a macro-scale level to choose the mechanical and geometrical characteristics of all layers is used. Each material layer is considered as homogeneous and elastic. The speed of the loads and the viscoelastic effects of eventual bituminous layers are taken into account indirectly by means of its equivalent elastic modulus. Several fracture mechanic concepts can be easily used. The solution method of M4-5n is implemented in the free Scilab software. The user of the resulting tool can calculate the finite values of mechanical fields in this multilayer structure (pavement) easily and quickly especially at the interfaces and at the crack edges without any problems of singularity. The version of this final tool, presented in this paper, is limited to study 2D cases only.

## 3 Study of a 2D composite pavement structure

The composite pavement structure studied here is chosen according to the French catalogue for pavements [32]. This type of structure is subjected to a heavy traffic type TC7 (between 17.5 and 43.5 million Trucks during all the life of the pavement). The pavement structure is composed of three layers and rests on a PF3 soil type. In the M4-5nW, we suppose that each M4-5n layer has the "real" thickness of the pavement and that the shear layer (M4-5n fourth layer) and the springs have the same Young modulus (Figure 1). All the mechanical and geometrical properties of the pavement and the soil are given below in Table 1.

Table 1: Geometrical and mechanical characteristics of the three layer pavement structure and the soil

| Material | Thickness (m) | Young's modulus (MPa) | Poisson's ratio |
|---|---|---|---|
| BBSG | $e^1=0.08$ | $E^1=5400$ | $\upsilon^1=0.35$ |
| GB3 | $e^2=0.15$ | $E^2=9300$ | $\upsilon^2=0.35$ |
| GC3 | $e^3=0.23$ | $E^3=23000$ | $\upsilon^3=0.25$ |
| Soil | - | $E^s=120$ | $\upsilon^4=0.35$ |

Where BBSG is a classical French bituminous mixes (NF EN 13108-1) such as GB3 for class-3 asphalt-treated roadbase aggregate (NF EN 13108-1) and GC3 is cement bound granular mixtures (NF EN 14227-1). Eventually, due to a shrinkage phenomenon that may occur in the third layer made of material treated by hydraulic cement, we assumed that this layer is cracked vertically across its thickness. In the case 2D plane strain, a unit pressure load (1 MPa) is chosen arbitrarily and uniformly distributed over a width (b-a) of 0.15m (Figure 1). We assume that the boundary conditions of blocked edges applied at the edges of the structure (Eq. 10) correspond to the confinement assumptions of the materials.

To choose the dimensions of the structure from which the load has no effect, a parametric study is done on an example of non-cracked pavement. The M4-5nW tool developed and programmed in Scilab is used. The results of M4-5nW calculation (the z axis upwards) are validated by comparison to those obtained by the Finite Element Method (FEM) given by the César-LCPC code (http://www.itech-soft.com/cesar/) (Figure 3). The 2D FEM mesh is made realized with quadratic assumptions and with help of the rectangular elements (Q8). A layer of 6m thick is usually admitted sufficient to represent the soil by FEM [40]. In the case of this example of non-cracked M4-5nW pavement structure, the zone III does not exist in the mesh (Figure 2). The thickness of the fourth layer is arbitrary chosen equal to $4e^3$. For both calculations of tools, there is no mesh optimization. We consider a regular mesh of a 0.001m width along the x-axis.

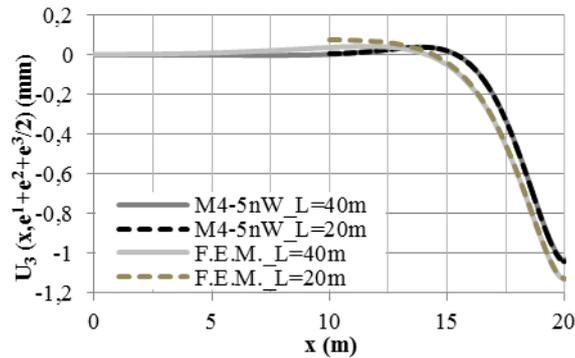

Figure 3 : Comparison of mean vertical displacement of layer 3 for different models

After various simulations, a 20m length for the M4-5nW pavement seems enough to ensure that the maximum deflection under load converges and that the vertical displacements of the pavement at the edges are zero (Figure 3). In the case of the FEM calculus, although the maximum deflection under load (greater than that obtained by the M4-5nW by 2.4%) converges to a length of 20m, it needs a 40m length for ensuring a zero displacement at the edges of the pavement (Figure 3).

In the following, we consider that a pavement of a 20m length is sufficient to make the comparison between models and validate the results obtained by M4-5nW simulation. The thickness $e^4$ of the shear layer is to be determined in the next paragraph.

## 3.1 Determination of the thickness of M45nW shear layer – case of 2D non-cracked pavement

In the case of the composite pavement of 20m long in the simulations, the M4-5nW assumptions (see section 2.2) indicate a maximum thickness of the shear layer of 0.92 m ($e^4_{max}=4e^3$). The influence of this thickness on the structural behaviour is carried out for four different values ($e^4 = e^3$, $2e^3$, $3e^3$ and $4e^3$) and four fields: for each layer i, the horizontal average displacement ($U_1^i(x)$) and the vertical average displacement ($U_3^i(x)$); at the interface i, i + 1, the shear interface stress ($\tau_1^{i,i+1}(x) = \sigma_{xz}(x, \sum_{i=1}^{i} e^i)$) and normal interface stress ($v^{i,i+1}(x) = \sigma_{zz}(x, \sum_{i=1}^{i} e^i)$).

For example, for the non-cracked composite pavement and different values of $e^4$, the vertical displacement upper of the first layer of the pavement and the interface normal stresses between the three layer pavement and the shear layer are shown in [34]. These values are compared with FEM results. Both simulations have the same mesh as for previous cases. This study leads to the following conclusions:
– The thickness $e^4$ of the shear layer has no influence on the horizontal displacements of layers ($U_1^i(x)$) or on shear interface stress and normal interface stress for the first two interfaces ($\tau_1^{1,2}(x)$, $\tau_1^{2,3}(x)$, $v^{1,2}(x)$, $v^{2,3}(x)$).
– It affects the vertical displacement of all the layers ($U_3^i(x)$) and the interface stresses of the third interface (between the pavement and the soil shear layer). The values of $U_3^i(x)$, $\tau_1^{3,4}(x)$ and $v^{3,4}(x)$ increase with the thickness of $e^4$.

For the interface shear stress of M4-5nW, the area under the curve gives a value equal to that obtained by FEM with a thickness $e^4=4e^3$.

## 4 M4-5nW analyses in the case of the 2D cracked pavement structure

Thereafter, in the case of the composite pavement structure studied in this paper, we suppose that the $e^4$ thickness of the shear layer is of 0.92m ($e^4_{max}= 0.92m$). Moreover, the third layer of the pavement generates a vertical crack, due to the shrinkage of the material treated with hydraulic binders. This crack either goes into the second layer (the phenomenon of "reflective cracking"); or contributes to a separation of the interface between the layers 2 and 3 [41-43]. In order to study this case, a vertical crack of 0.001m width is introduced across the layer 3. According to the principles of linear elastic fracture mechanics that are easily used in this type of modelling [26], this initial crack modifies only the boundary conditions of the concerned layer of the structure. Subsequently, it is assumed that the load is positioned on the left of the crack.

In this case, an irregular mesh is chosen in the zones I and IV (see Figure 2) in order to optimize computation time and obtain accurate M4-5n intensities of the interface stresses under load and at the edge of the crack. These interface fields may be used to understand the interface failure mechanisms of the initial cracked pavement structure.

## 4.1 Validation of the M4-5nW vs the FEM

In the case of a load located on the centre of the pavement and on the left of a vertical crack edge (positioned in the layer 3 between the points x= 10.075m and x= 10.076m), a parametric study of the convergence of interface stress fields of the M4-5n (between the layers 2 and 3) is done. To realize this study, first we consider that the length of the mesh in zone II under the load is equal to the length of the first mesh in zone IV just after the crack. According to the formula used to determine the distribution of mesh in different zones, we can determine the number of nodes in zone II and IV. And then, the number of nodes in zone I is adjusted in order to obtain a length of the last mesh equal to those in zone II. Different values are tested and it is then found that 74 nodes in zone I, 121 nodes in zone II (under the load) and 70 nodes in zone IV are sufficient to obtain finite intensities with a precision of $10^{-2}$ MPa at the left and right tip of the crack. The length of the mesh is then 0.00125mm. Zone III contains 2 nodes to characterize the crack (Figure 2). For example Figure 4 shows the variation of the intensity of normal stress along the interface 2,3 ($v^{2,3}(x)$) at the left tip of the crack for x=b=10.075m (a) and the right tip of the crack for x=10.076 (b), as function of the number of nodes in zone II and IV respectively.

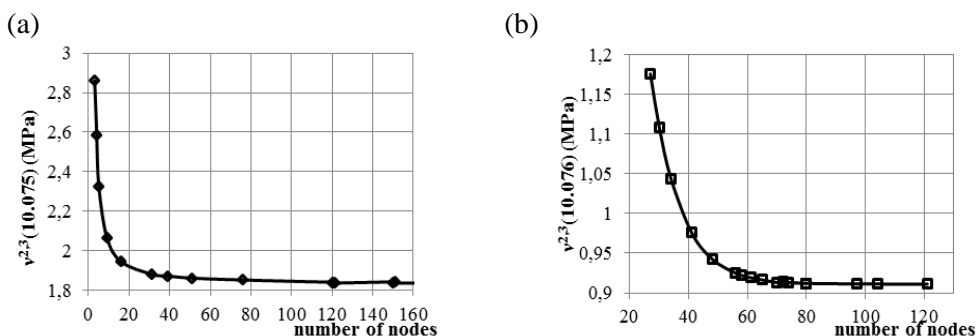

Figure 4: Converge of interface normal stress between layers 2 and 3 at: (a) the left tip of the crack; (b) the right tip of the crack

Figure 5 thus illustrates a comparison between simulations made by FEM (finite elements Q8) and M4-5nW results regarding the shear stress (a) and normal stress (b) at the interface between the layers 3 and 4.

The results are normalized by the maximum of stresses obtained by the FEM ($\sigma_{xz}^M$ and $\sigma_{zz}^M$). We note that the simulations of the shear stress and the normal stress at the interface are consistent for both models excluding their intensity at the crack edge as expected (the area under the curves representing the interface shear and normal stress between layers 3 and 4 given by the M4-5nW is lower than that given by the FEM by 0.7% and 0. 1% respectively).

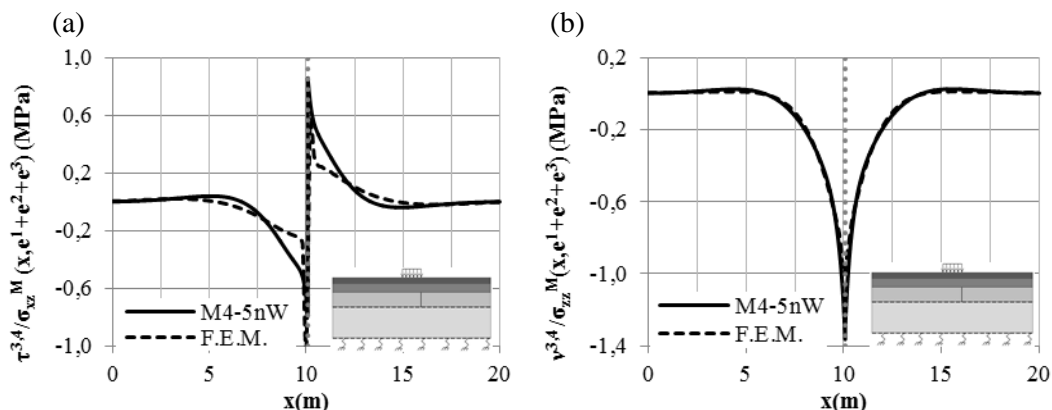

Figure 5 : The M4-5nW results compared to the FEM ones (a) interface shear stress between layers 3 and 4; (b) interface normal stress between layers 3 and 4

In the Table 2 below, the time of calculation CPU obtained for the two models used here is presented for the case of 2D plane strain. We note that for the same number of regular meshes, the time obtained for M4-5nW is very interesting compared to the one obtained by FEM (6 times faster). This result responds to the objective of building a fast calculation tool in order to use it in parametric studies.

Table 2: Comparison of time of calculation CPU between the two models in the cracked pavement case

|  | M4-5nW | F.E.M. |
|---|---|---|
| Time CPU (s) | 0.89 | 5.4 |

These results show clearly the interest of M4-5nW thus built. In fact, the proposed approach reducing of one dimension the studied problem needs indeed for its development a high number of unknowns and equations. But this difficulty is hidden for the users of the "M4-5nW" tool, in which this model is implemented. As said before, the real object 3D (2D) is transforming into one geometry 2D (1D) which makes the mesh less complex and reduces the number of elements, then the calculus time. The developed method using the M4-5n is implemented in a tool "Scilab" which is easy to use to calculate the finite values of mechanical fields in a multilayer structure (pavement) especially at the interfaces and at the crack edges without problems of singularity. This tool also allows practical parametric studies. All these elements represent the advantages of the model developed in this paper, which becomes more interesting against classical FEM used when the tool is developed to study 3D cases [45-48].

## 4.2 Evaluation of interface stresses of the M4-5nW for a cracked composite pavement subjected to different positions of load

In order to illustrate the ability of the pavement to generate delamination at the interface between layer 2 and layer 3 during the reflective cracking phenomenon, different load case positions are studied to understand the influence of its position with respect to the vertical crack location in the third layer. The M4-5nW developed is used. Figures 6 and 7 present respectively the interface shear $\tau^{2,3}(x)$ and normal stress $v^{2,3}(x)$ between layers 2 and 3 for four cases of pavement calculations: load at the centre of non-cracked pavement (case a); load position at 1m to the left of the crack located in the third layer (case b1); load position at 0.5m to the left of the crack (case b2); load at the left edge of crack (case b3).

First, due to the fact that the M4-5nW intensity of the shear stress and normal stress at the interface are finite near the crack, we can compare the different results between themselves. For the case of non-cracked pavement (case a), we note that the curve of the shear stress and the normal stress are symmetrical at the crack location. The maximum of the value of these stresses is low. For the case of the cracked pavement structure, the maximum of the intensity of the shear stress and the normal stress near the crack at the interface 2,3 (just above the crack) is increasing when the load is moving to the crack. It reaches its maximum value for the case b3, i.e. when the load is at edge of the crack. Their maximum intensity is quite comparable. If the interface resistance is smaller than the layer 2 material (that is working under tension condition), a debonding phenomenon may happen in a mixed mode fracture condition. This phenomenon may occur either under heavy loading or per fatigue due to the traffic of the loads. In addition, along the moving load course, on this interface a non-negligible shear stress value exists before the crack area (Figure 6).

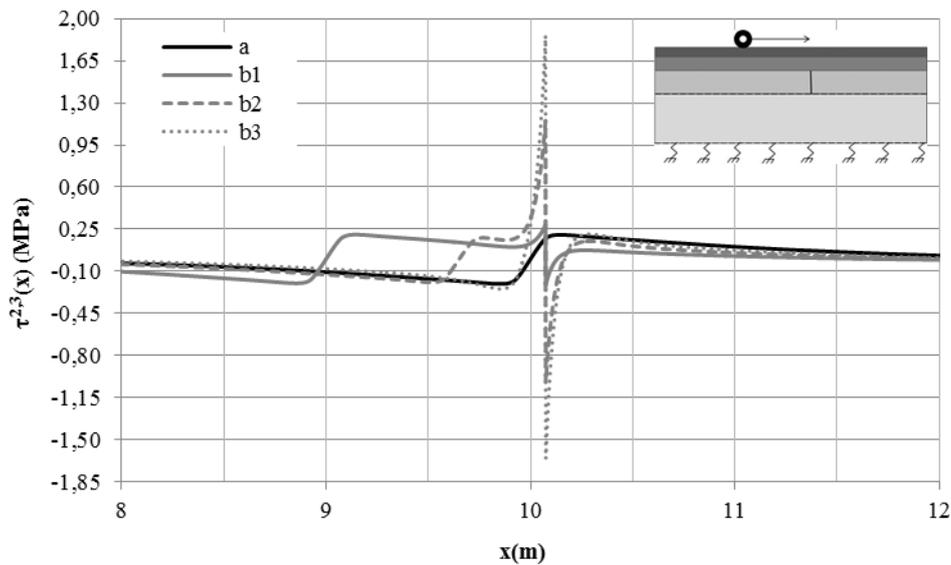

Figure 6 : A comparison of the distribution of shear stresses of the M4-5n at the interface 2,3 for four loading position cases (a, b1, b2 and b3)

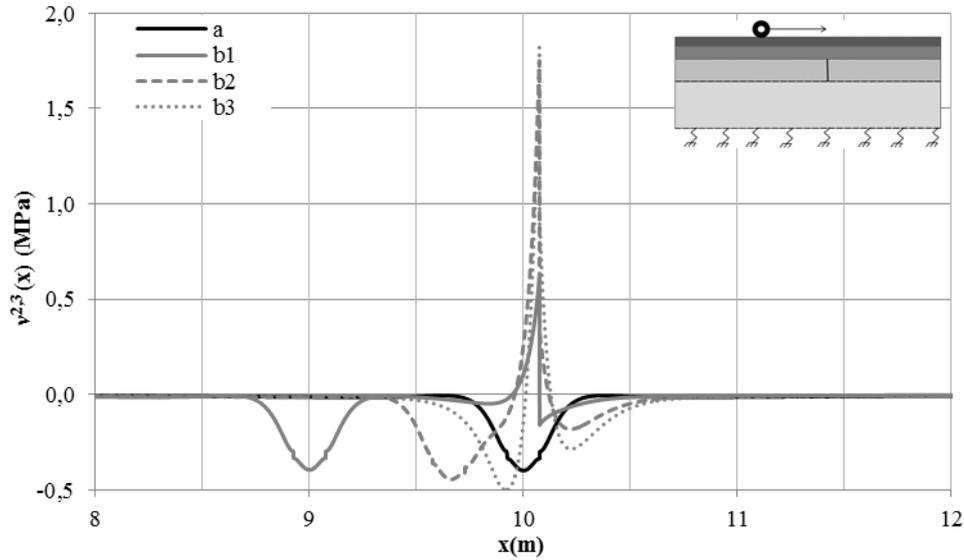

Figure 7: A comparison of the distribution of normal stress of the M4-5n at the interface 2,3 for four loading position cases (a, b1, b2 and b3)

With the help of such a M4-5nW tool, this short analysis confirms that the position of the load with respect to the crack location plays an important role in the analysis of the behaviour of the 2D cracked pavement structure. The risk of interface damage of a pavement structure having a vertical crack is higher when the load is on the edge of this crack than for other loading positions. It is sufficient to consider this case in any 2D analysis of pavement with a single vertical crack across a layer.

## 4.3 Evaluation of interface stresses of the M4-5nW for a composite pavement subjected to temperature changing

In the case of the cracked composite pavement previously studied with a load located at the edge of the vertical crack (b3 case of Figures 6 and 7), the temperature changing (between day and night) is studied on the values of the interface shear and normal stresses. The temperature influences the bituminous materials depending in which location the layers are placed with or not in contact with the outside temperature. Actually, according to some full-scale data [44], the second layer usually receives a thermal delay in the depth of the pavement. The temperature affects the stiffness of the bituminous materials that have thermos-susceptible behaviour properties. In case of a hot external temperature, the Young's modulus of the first layer decreases in comparison of its first value. We consider that the ratio of modulus between the first and second layer is equal to ¼, with $E^1$= 2325 MPa and $E^2$= 9300MPa (initially the ratio is ½ with $E^1$= 5400MPa and $E^2$= 9300MPa). On the contrary, in the case of a cold external temperature, the Young's modulus of the first layer increases and the case of a ratio of modulus between layers equal to 1 is taken, with $E^1$= 9300MPa and $E^2$= 9300MPa. Figures 8 and 9 illustrate a zoom of the interface shear stress $\tau^{2,3}(x)$ and normal stress $v^{2,3}(x)$ between layers 2 and 3 respectively for the three ratio of modulus between layers 1 and 2: $E^1/E^2$=1/2, $E^1/E^2=1/4$ and $E^1/E^2=1$. We note that the different values of the Young modulus of the first layer of the composite pavement cracked in its third layer have few effects on the intensity of the M4-5nW interface stresses between layer 2,3 especially for the shear stress (Figure 8). These effects are not so much huge as those obtained previously in the case of the moving load at several positions. Nethertheless, cumulated to the moving load, it may favor reflective cracking phenomenon in the structure.

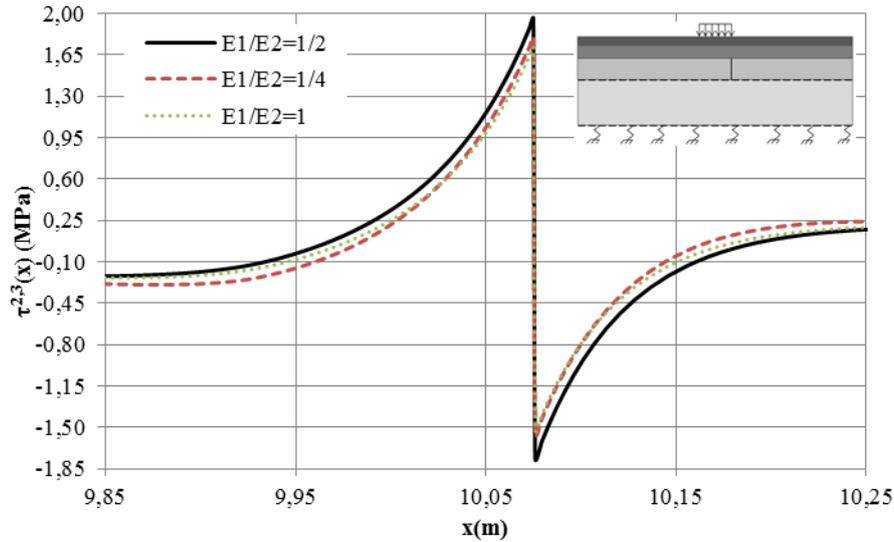

Figure 8: Thermal variation effect in the 1st bituminous material layer on the M4-5nW shear stress distribution at the interface 2,3 of a composite cracked pavement

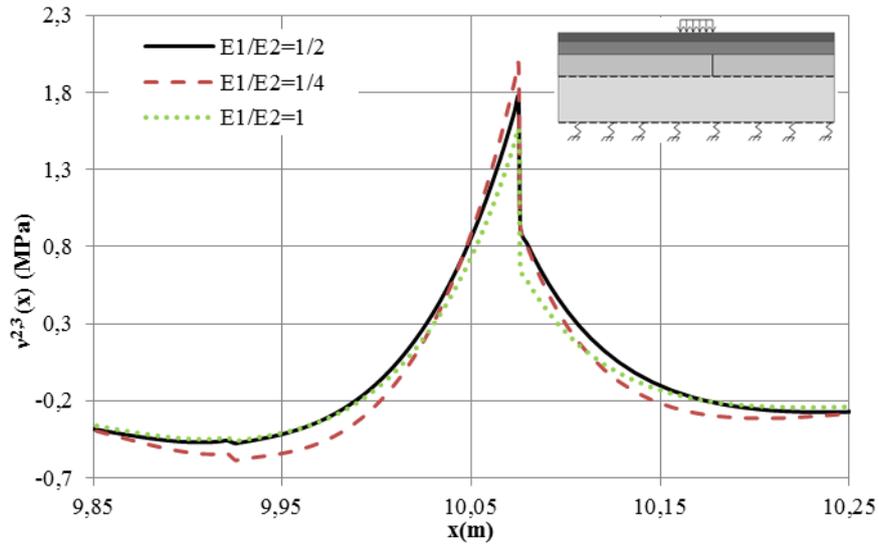

Figure 9: Thermal variation effect in the 1st bituminous material layer on the M4-5nW normal stress distribution at the interface 2,3 of a composite cracked pavement

# 5 Conclusions and perspectives

In this paper, we propose to model the pavement structures with initial vertical crack across one layer with the help of a simplified modelling tool. The tool is based on the M4-5n that is a model of multi-layer structures dedicated to a simplified bending analysis of delamination problems in composite structures [21-23] [25]. On the contrary to classical FEM tool, in the proposed approach the problem reduction of 1 dimension and the regular mechanical stress fields at the interface and vertical crack locations offer practical parametric studies for pavements with such discontinuities. In this half analytical elastic proposed solution dedicated to the analysis of pavements that have such strong discontinuities, the speed of the loads and the viscoelastic effects of eventual bituminous layers are taken into account indirectly by means of its equivalent elastic modulus. The pavement structure is chosen to be equivalent to three elastic layers resting on a soil. The soil is modelled by a massive of Winkler (W) and an additional shear layer to ensure the transfer of shear stresses between the pavement and the soil. It leads to practical and efficient calculations for the analysis of fracture behaviour of pavement at a real-scale. The pavement and the soil constitute the M4-5nW. All the layers are modelled by the M4-5n. This new approach of the pavement modelling is then called the M4-5nW.

In the case of 2D plane strain, all the equations of the M4-5n are written analytically. These second order differential equations are solved by the finite difference method (Newmark). The numerical solution of the system of equations is validated, by comparing the results of the M4-5n with those obtained by a finite element code, with a very promising CPU time. This time saving is due to the choice of the modelling for the soil and the use of the M4-5n that reduces the dimension of the studied problem. Thus the real object 3D (2D) is transforming into one geometry 2D (1D). Furthermore, the M4 modelling approach avoids singularities by giving a finite value of stresses at plate edges. Between two adjacent material layers, it becomes possible to express a delamination criterion in terms of interfacial stresses [21] [25-26] [36]. The time saving is also due to the simplified writing of matrixes needed to solve the problem as well as the optimum number of meshes chosen. Although the number of M4-5n unknowns to be solved is quite large, the semi analytical solution of the equations of this model (for the finite difference method) allows practical parametric studies. This computational simple tool aims to be used by the engineers especially for the macro-scale analysis of cracked pavement structures. It provides reference solutions for further 3D developments of a more general engineering software [45].

The mechanical response of a 3D cracked structure, representative of a flexible pavement tested under full-scale conditions during an accelerated fatigue test performed at IFSTTAR is also possible. Recently, first scenarios of cracking have been successfully computed and proposed [46-48]. A partial debonding of interface between layers will be introduced in the modelling in order to deep those results and other full-scale experimental data [49-50]. In addition, according to previous works [28-29], thermal gradients will be introduced in such developments in order to study the degradation of cement concrete pavements and the durability of next road concepts of the future.

# Appendix A

Let's note

$$\lambda^i = \frac{e^i(1-\nu^i)}{E^i}, \kappa^i = \frac{e^i}{E^i} \quad \text{Eq. (A.1)}$$

$$\Delta 1 = \left(\frac{\lambda^3}{12}\right)^2 \left(\frac{\lambda^1}{4}+\frac{\lambda^2}{4}\right) + \left(\frac{\lambda^2}{12}\right)^2 \left(\frac{\lambda^3}{4}+\frac{\lambda^4}{4}\right) - \left(\frac{\lambda^1}{4}+\frac{\lambda^2}{4}\right)\left(\frac{\lambda^2}{4}+\frac{\lambda^3}{4}\right)\left(\frac{\lambda^3}{4}+\frac{\lambda^4}{4}\right) \quad \text{Eq. (A.2)}$$

$$\Delta 2 = 13\left(\left(\frac{9}{2}\frac{\kappa^3}{35}\right)^2 \left(\frac{\kappa^1}{35}+\frac{\kappa^2}{35}\right) + \left(\left(\frac{9}{2}\frac{\kappa^2}{35}\right)^2 - 13^2 \left(\frac{\kappa^1}{35}+\frac{\kappa^2}{35}\right)\left(\frac{\kappa^2}{35}+\frac{\kappa^3}{35}\right)\right)\left(\frac{\kappa^3}{35}+\frac{\kappa^4}{35}\right)\right) \quad \text{Eq. (A.3)}$$

and let's notice that all coefficients written in the following matrixes A, B and C are proportional to the quotients $1/\kappa^i$ that is to say $\frac{E^i}{e^i}$ and to 1 in the matrixes D, E, F and G.

**Matrix A**

$$A = \begin{pmatrix}
\frac{e^1 E^1}{1-\nu^{1^2}} & 0 & 0 & 0 & 0 & 0 & 0 & 0 & 0 & 0 & 0 & 0 \\
0 & \frac{e^{1^2} E^1}{5(1-\nu^{1^2})} & 0 & 0 & 0 & 0 & 0 & 0 & 0 & 0 & 0 & 0 \\
0 & 0 & e^1 A_{33} & 0 & 0 & e^2 A_{36} & 0 & 0 & -e^3 A_{39} & 0 & 0 & -e^4 A_{312} \\
0 & 0 & 0 & \frac{e^2 E^2}{1-\nu^{2^2}} & 0 & 0 & 0 & 0 & 0 & 0 & 0 & 0 \\
0 & 0 & 0 & 0 & \frac{e^{2^2} E^2}{5(1-\nu^{2^2})} & 0 & 0 & 0 & 0 & 0 & 0 & 0 \\
0 & 0 & e^1 A_{36} & 0 & 0 & e^2 A_{66} & 0 & 0 & -e^3 A_{69} & 0 & 0 & -e^4 A_{612} \\
0 & 0 & 0 & 0 & 0 & 0 & \frac{e^3 E^3}{1-\nu^{3^2}} & 0 & 0 & 0 & 0 & 0 \\
0 & 0 & 0 & 0 & 0 & 0 & 0 & \frac{e^{3^2} E^3}{5(1-\nu^{3^2})} & 0 & 0 & 0 & 0 \\
0 & 0 & -e^1 A_{39} & 0 & 0 & -e^2 A_{69} & 0 & 0 & e^3 A_{99} & 0 & 0 & e^4 A_{912} \\
0 & 0 & 0 & 0 & 0 & 0 & 0 & 0 & 0 & \frac{e^4 E^4}{1-\nu^{4^2}} & 0 & 0 \\
0 & 0 & 0 & 0 & 0 & 0 & 0 & 0 & 0 & 0 & \frac{e^{4^2} E^4}{5(1-\nu^{4^2})} & 0 \\
0 & 0 & -e^1 A_{312} & 0 & 0 & -e^2 A_{612} & 0 & 0 & e^3 A_{912} & 0 & 0 & e^4 A_{1212}
\end{pmatrix}$$

$$A_{33} = \frac{5E^1}{e^1(1+\nu^1)} + a_{33} \quad , \quad A_{66} = \frac{5E^2}{e^2(1+\nu^2)} + a_{66}$$

$$A_{99} = \frac{5E^3}{e^3(1+\nu^3)} + a_{99} \quad , \quad A_{1212} = \frac{5E^4}{e^4(1+\nu^4)} + a_{1212} \quad \text{Eq. (A.4)}$$

With the following 6 $A_{jk}$ and 4 $a_{jk}$ coefficients

$$a_{33} = \frac{1}{12\Delta 1}\left(\left(\frac{\lambda^3}{12}\right)^2 - \left(\frac{\lambda^2}{4}+\frac{\lambda^3}{4}\right)\left(\frac{\lambda^3}{4}+\frac{\lambda^4}{4}\right)\right) \; ; \; A_{36} = \frac{1}{12\Delta 1}\left(\left(\frac{\lambda^3}{12}\right)^2 - \left(\frac{\lambda^2}{3}+\frac{\lambda^3}{4}\right)\left(\frac{\lambda^3}{4}+\frac{\lambda^4}{4}\right)\right) \quad \text{Eq. (A.5)}$$

$$A_{39} = \frac{\lambda^2}{12^2 \Delta 1}\left(\frac{\lambda^3}{3}+\frac{\lambda^4}{4}\right); \; A_{312} = \frac{\lambda^2 \lambda^3}{12^3 \Delta 1}; \; a_{66} = \frac{1}{12\Delta 1}\left(\left(\frac{\lambda^3}{12}\right)^2 - \left(\frac{\lambda^1}{4}+\frac{2\lambda^2}{3}+\frac{\lambda^3}{4}\right)\left(\frac{\lambda^3}{4}+\frac{\lambda^4}{4}\right)\right)$$

$$A_{69} = \frac{1}{12\Delta 1}\left(\frac{\lambda^1}{4}+\frac{\lambda^2}{3}\right)\left(\frac{\lambda^3}{3}+\frac{\lambda^4}{4}\right); \; A_{612} = \frac{\lambda^3}{12^2 \Delta 1}\left(\frac{\lambda^1}{4}+\frac{\lambda^2}{3}\right)$$

$$a_{99} = \frac{1}{12\Delta 1}\left(\left(\frac{\lambda^2}{12}\right)^2 - \left(\frac{\lambda^1}{4}+\frac{\lambda^2}{4}\right)\left(\frac{\lambda^2}{4}+\frac{2\lambda^3}{3}+\frac{\lambda^4}{4}\right)\right)$$

$$A_{912} = \frac{1}{12\Delta 1}\left(\left(\frac{\lambda^2}{12}\right)^2 - \left(\frac{\lambda^1}{4}+\frac{\lambda^2}{4}\right)\left(\frac{\lambda^2}{4}+\frac{\lambda^3}{3}\right)\right); \; a_{1212} = \frac{1}{12\Delta 1}\left(\left(\frac{\lambda^2}{12}\right)^2 - \left(\frac{\lambda^1}{4}+\frac{\lambda^2}{4}\right)\left(\frac{\lambda^2}{4}+\frac{\lambda^3}{4}\right)\right)$$

**Matrix B**

$$B = \begin{pmatrix}
0 & 0 & \frac{e^1}{12}B_{13} & 0 & 0 & \frac{e^2}{12}B_{16} & 0 & 0 & \frac{-e^3}{12}B_{19} & 0 & 0 & \frac{-e^4}{12}B_{112} \\
0 & 0 & e^1 B_{23} & 0 & 0 & \frac{e^2}{12}B_{16} & 0 & 0 & \frac{-e^3}{12}B_{19} & 0 & 0 & \frac{-e^4}{12}B_{112} \\
-B_{13} & -\frac{5e^1}{12}B_{23} & 0 & B_{34} & -\frac{5e^2}{12}B_{16} & 0 & -B_{37} & \frac{5e^3}{12}B_{19} & 0 & -B_{112} & \frac{5e^4}{12}B_{112} & 0 \\
0 & 0 & -\frac{e^1}{12}B_{34} & 0 & 0 & -\frac{e^2}{12}B_{46} & 0 & 0 & \frac{-e^3}{12}B_{49} & 0 & 0 & \frac{-e^4}{12}B_{412} \\
0 & 0 & \frac{e^1}{12}B_{16} & 0 & 0 & \frac{e^2}{12}B_{56} & 0 & 0 & \frac{-e^3}{12}B_{59} & 0 & 0 & \frac{-e^4}{12}B_{512} \\
-B_{16} & -\frac{5e^1}{12}B_{16} & 0 & B_{46} & -\frac{5e^2}{12}B_{56} & 0 & -B_{67} & \frac{5e^3}{12}B_{59} & 0 & -B_{512} & \frac{5e^4}{12}B_{512} & 0 \\
0 & 0 & \frac{e^1}{12}B_{37} & 0 & 0 & \frac{e^2}{12}B_{67} & 0 & 0 & \frac{e^3}{12}B_{79} & 0 & 0 & \frac{e^4}{12}B_{712} \\
0 & 0 & -\frac{e^1}{12}B_{19} & 0 & 0 & -\frac{e^2}{12}B_{59} & 0 & 0 & \frac{e^3}{12}B_{89} & 0 & 0 & \frac{e^4}{12}B_{812} \\
B_{19} & \frac{5e^1}{12}B_{19} & 0 & B_{49} & \frac{5e^2}{12}B_{59} & 0 & -B_{79} & -\frac{5e^3}{12}B_{89} & 0 & B_{812} & \frac{-5e^4}{12}B_{812} & 0 \\
0 & 0 & \frac{e^1}{12}B_{112} & 0 & 0 & \frac{e^2}{12}B_{512} & 0 & 0 & \frac{-e^3}{12}B_{812} & 0 & 0 & \frac{-e^4}{12}B_{1012} \\
0 & 0 & -\frac{e^1}{12}B_{112} & 0 & 0 & -\frac{e^2}{12}B_{512} & 0 & 0 & \frac{e^3}{12}B_{812} & 0 & 0 & \frac{e^4}{12}B_{1112} \\
B_{112} & \frac{5e^1}{12}B_{112} & 0 & B_{412} & \frac{5e^2}{12}B_{512} & 0 & -B_{712} & \frac{-5e^3}{12}B_{812} & 0 & B_{1012} & \frac{-5e^4}{12}B_{1112} & 0
\end{pmatrix}$$

$$B_{23} = \frac{-12E^1}{e^1(1+\nu^1)} + B_{13} \;,\; B_{56} = \frac{-12 E^2}{e^2(1+\nu^2)} + b_{56}$$
$$B_{89} = \frac{-12E^3}{e^3(1+\nu^3)} + b_{89} \;,\; B_{1112} = \frac{-12E^4}{e^4(1+\nu^4)} + B_{1012}$$

Eq. (A.6)

With the following 16 $B_{jk}$ and 2 $b_{jk}$ coefficients (some of them are linked to the previous $A_{jk}$ ones)

$$B_{13} = 12a_{33} \;;\; B_{16} = 12A_{36};\; B_{19} = 12A_{39} \;;\; B_{112} = 12A_{312}$$
$$B_{34} = \frac{1}{\Delta 1}\left(\left(\frac{\lambda^3}{12}\right)^2 - \left(\frac{\lambda^2}{6}+\frac{\lambda^3}{4}\right)\left(\frac{\lambda^3}{4}+\frac{\lambda^4}{4}\right)\right);\; B_{37} = \frac{\lambda^2}{12\Delta 1}\left(\frac{\lambda^3}{6}+\frac{\lambda^4}{4}\right)$$
$$B_{46} = \frac{1}{\Delta 1}\left(\left(\frac{\lambda^3}{12}\right)^2 - \left(-\frac{\lambda^1}{4}+\frac{\lambda^3}{4}\right)\left(\frac{\lambda^3}{4}+\frac{\lambda^4}{4}\right)\right);\; B_{49} = \frac{1}{\Delta 1}\left(\frac{\lambda^1}{4}+\frac{\lambda^2}{6}\right)\left(\frac{\lambda^3}{3}+\frac{\lambda^4}{4}\right)$$
$$B_{412} = \frac{\lambda^3}{12\Delta 1}\left(\frac{\lambda^1}{4}+\frac{\lambda^2}{6}\right);\; b_{56} = 12a_{66} \;;\; B_{59} = 12A_{69} \;;\; B_{512} = 12A_{612}$$
$$B_{67} = \frac{1}{\Delta 1}\left(\frac{\lambda^1}{4}+\frac{\lambda^2}{3}\right)\left(\frac{\lambda^3}{6}+\frac{\lambda^4}{4}\right) \;;\; B_{79} = \frac{1}{\Delta 1}\left(\left(\frac{\lambda^2}{12}\right)^2 + \left(\frac{\lambda^1}{3}+\frac{\lambda^2}{4}\right)\left(-\frac{\lambda^2}{4}+\frac{\lambda^4}{4}\right)\right)$$
$$B_{712} = \frac{1}{\Delta 1}\left(\left(\frac{\lambda^2}{12}\right)^2 - \left(\frac{\lambda^1}{4}+\frac{\lambda^2}{4}\right)\left(\frac{\lambda^2}{4}+\frac{\lambda^3}{6}\right)\right);\; b_{89} = 12a_{99} \;;\; B_{812} = 12A_{912};\; B_{1012} =$$
$$12a_{1212}$$

Eq. (A.7)

**Matrix C**

$$C = \begin{pmatrix} -C_{11} & \frac{-5e^1}{12}C_{11} & 0 & C_{14} & \frac{-5e^2}{12}C_{15} & 0 & -C_{17} & \frac{5e^3}{12}C_{18} & 0 & -C_{110} & \frac{5e^4}{12}C_{110} & 0 \\ -C_{11} & \frac{-5e^1}{12}C_{22} & 0 & C_{14} & \frac{-5e^2}{12}C_{15} & 0 & -C_{17} & \frac{5e^3}{12}C_{18} & 0 & -C_{110} & \frac{5e^4}{12}C_{110} & 0 \\ 0 & 0 & \frac{-12}{e^1}C_{33} & 0 & 0 & \frac{12}{e^1}C_{36} & 0 & 0 & \frac{12}{e^1}C_{39} & 0 & 0 & \frac{-12}{e^1}C_{312} \\ C_{14} & \frac{5e^1}{12}C_{14} & 0 & -C_{44} & \frac{5e^2}{12}C_{45} & 0 & -C_{47} & \frac{5e^3}{12}C_{48} & 0 & -C_{410} & \frac{5e^4}{12}C_{410} & 0 \\ C_{15} & \frac{-5e^1}{12}C_{15} & 0 & C_{45} & -\frac{5e^2}{12}C_{55} & 0 & -C_{57} & \frac{5e^3}{12}C_{58} & 0 & -C_{510} & \frac{5e^4}{12}C_{510} & 0 \\ 0 & 0 & \frac{12}{e^2}C_{36} & 0 & 0 & \frac{-12}{e^2}C_{66} & 0 & 0 & \frac{-12}{e^2}C_{69} & 0 & 0 & \frac{12}{e^2}C_{612} \\ -C_{17} & \frac{-5e^1}{12}C_{17} & 0 & -C_{47} & \frac{-5e^2}{12}C_{57} & 0 & -C_{77} & \frac{-5e^3}{12}C_{78} & 0 & C_{710} & \frac{-5e^4}{12}C_{710} & 0 \\ C_{18} & \frac{5e^1}{12}C_{18} & 0 & C_{48} & \frac{5e^2}{12}C_{58} & 0 & -C_{78} & \frac{-5e^3}{12}C_{88} & 0 & C_{810} & \frac{-5e^4}{12}C_{810} & 0 \\ 0 & 0 & \frac{12}{e^3}C_{39} & 0 & 0 & \frac{-12}{e^3}C_{69} & 0 & 0 & \frac{-12}{e^3}C_{99} & 0 & 0 & \frac{12}{e^3}C_{912} \\ -C_{110} & \frac{-5e^1}{12}C_{110} & 0 & -C_{410} & \frac{-5e^2}{12}C_{510} & 0 & C_{710} & \frac{5e^3}{12}C_{810} & 0 & -C_{1010} & \frac{5e^4}{12}C_{1010} & 0 \\ C_{110} & \frac{5e^1}{12}C_{110} & 0 & C_{410} & \frac{5e^2}{12}C_{510} & 0 & -C_{711} & \frac{-5e^3}{12}C_{810} & 0 & C_{1010} & \frac{-5e^4}{12}C_{1010} & 0 \\ 0 & 0 & \frac{-12}{e^4}C_{312} & 0 & 0 & \frac{12}{e^4}C_{612} & 0 & 0 & \frac{12}{e^4}C_{912} & 0 & 0 & \frac{-12}{e^4}C_{1212} \end{pmatrix}$$

$$C_{22} = \frac{12E^1}{5e^1(1+v^1)} + C_{11}, \quad C_{55} = \frac{12E^2}{5e^2(1+v^2)} + c_{55}$$

$$C_{88} = \frac{12\,E^3}{5e^3(1+v^3)} + c_{88}, \quad C_{1111} = \frac{12E^4}{5e^4(1+v^4)} + C_{1010}$$

Eq. (A.8)

The 30 $C_{jk}$ and 2 $c_{jk}$ coefficients are proportional to the quotients: $\frac{E^i}{e^i}$ (some of them are linked to the previous $A_{jk}$ and $B_{jk}$ ones).

$$C_{11} = 12a_{33};\ C_{14} = B_{34};\ C_{15} = 12A_{36};\ C_{17} = B_{37};\ C_{18} = 12A_{39};\ C_{110} = 12A_{312}$$

$$C_{33} = \frac{1}{\Delta 2}\left(\left(\frac{9}{2}\frac{\kappa^3}{35}\right)^2 - 13^2\left(\frac{\kappa^2}{35} + \frac{\kappa^3}{35}\right)\left(\frac{\kappa^3}{35} + \frac{\kappa^4}{35}\right)\right)$$

$$C_{36} = \frac{1}{\Delta 2}\left(\left(\frac{9}{2}\frac{\kappa^3}{35}\right)^2 - 13\left(\frac{\kappa^2}{2} + \frac{13\kappa^3}{35}\right)\left(\frac{\kappa^3}{35} + \frac{\kappa^4}{35}\right)\right);\ C_{39} = \frac{1}{\Delta 2}\left(\frac{9}{2}\frac{\kappa^2}{35}\right)\left(\frac{\kappa^3}{2} + \frac{13\kappa^4}{35}\right)$$

$$C_{312} = \frac{1}{\Delta 2}\left(\frac{9}{2}\frac{\kappa^2}{35}\right)\left(\frac{9}{2}\frac{\kappa^3}{35}\right);\ C_{44} = \frac{1}{\Delta 1}\left(\left(\frac{\lambda^3}{12}\right)^2 - \left(\frac{\lambda^1}{4} + \frac{\lambda^2}{3} + \frac{\lambda^3}{4}\right)\left(\frac{\lambda^3}{4} + \frac{\lambda^4}{4}\right)\right);\ C_{45} = B_{46}$$

$$C_{47} = \frac{1}{\Delta 1}\left(\frac{\lambda^1}{4} + \frac{\lambda^2}{6}\right)\left(\frac{\lambda^3}{6} + \frac{\lambda^4}{4}\right);\ C_{48} = B_{49};\ C_{410} = B_{412}$$

$$c_{55} = 12a_{66};\ C_{57} = \frac{1}{\Delta 1}\left(\frac{\lambda^1}{4} + \frac{\lambda^2}{3}\right)\left(\frac{\lambda^3}{6} + \frac{\lambda^4}{4}\right);\ C_{58} = 12A_{69};\ C_{510} = 12A_{612}$$

$$C_{66} = \frac{1}{\Delta 2}\left(\left(\frac{9}{2}\frac{\kappa^3}{35}\right)^2 - 13\left(\frac{13\kappa^1}{35} + \kappa^2 + \frac{13\kappa^3}{35}\right)\left(\frac{\kappa^3}{35} + \frac{\kappa^4}{35}\right)\right);\ C_{69} = \frac{1}{\Delta 2}\left(\frac{13\kappa^1}{35} + \frac{\kappa^2}{2}\right)\left(\frac{\kappa^3}{2} + \frac{13\kappa^4}{35}\right)$$

Eq. (A.9)

$$C_{612} = \frac{1}{\Delta 2}\left(\frac{9}{2}\frac{\kappa^3}{35}\right)\left(\frac{13\kappa^1}{35} + \frac{\kappa^2}{2}\right);\ C_{77} = \frac{1}{\Delta 1}\left(\left(\frac{\lambda^2}{12}\right)^2 - \left(\frac{\lambda^1}{4} + \frac{\lambda^2}{4}\right)\left(\frac{\lambda^2}{4} + \frac{\lambda^3}{3} + \frac{\lambda^4}{4}\right)\right)$$

$$C_{78} = \frac{1}{\Delta 1}\left(\left(\frac{\lambda^2}{12}\right)^2 - \left(\frac{\lambda^1}{4} + \frac{\lambda^2}{4}\right)\left(\frac{\lambda^2}{4} - \frac{\lambda^4}{4}\right)\right);\ C_{710} = \frac{1}{\Delta 1}\left(\left(\frac{\lambda^2}{12}\right)^2 - \left(\frac{\lambda^1}{4} + \frac{\lambda^2}{4}\right)\left(\frac{\lambda^2}{4} + \frac{\lambda^3}{6}\right)\right)$$

$$c_{88} = 12a_{99};\ C_{810} = 12A_{912};\ C_{99} = \frac{1}{\Delta 2}\left(\left(\frac{9}{2}\frac{\kappa^2}{35}\right)^2 - 13\left(\frac{\kappa^1}{35} + \frac{\kappa^2}{35}\right)\left(\frac{13\kappa^2}{35} + \kappa^3 + \frac{13\kappa^4}{35}\right)\right)$$

$$C_{912} = \frac{1}{\Delta 2}\left(\left(\frac{9}{2}\frac{\kappa^2}{35}\right)^2 - 13\left(\frac{\kappa^1}{35} + \frac{\kappa^2}{35}\right)\left(\frac{13\kappa^2}{35} + \frac{\kappa^3}{2}\right)\right);\ C_{1010} = 12A_{1212}$$

$$C_{1212} = \frac{1}{\Delta 2}\left(\left(\frac{9}{2}\frac{\kappa^2}{35}\right)^2 - 13^2\left(\frac{\kappa^1}{35} + \frac{\kappa^2}{35}\right)\left(\frac{\kappa^2}{35} + \frac{\kappa^3}{35}\right)\right)$$

**Matrix D and E**

$$D=\begin{pmatrix} 0 & 0 \\ 0 & 0 \\ -1-D_{31} & 0 \\ 0 & 0 \\ 0 & 0 \\ -D_{61} & 0 \\ 0 & 0 \\ 0 & 0 \\ D_{91} & 0 \\ 0 & 0 \\ 0 & 0 \\ D_{121} & 0 \end{pmatrix} \quad E=\begin{pmatrix} 0 & 0 \\ 0 & 0 \\ E_{31} & 0 \\ 0 & 0 \\ 0 & 0 \\ E_{61} & 0 \\ 0 & 0 \\ 0 & 0 \\ -E_{91} & 0 \\ 0 & 0 \\ 0 & 0 \\ -1-E_{121} & 0 \end{pmatrix} \qquad \text{Eq. (A.10)}$$

The 4 $D_{jk}$ and $E_{jk}$ coefficients are:

$$D_{31} = \lambda^1 a_{11}; D_{61} = \lambda^1 A_{36}; D_{91} = \lambda^1 A_{39}; D_{121} = \lambda^1 A_{312}$$
$$E_{31} = \lambda^4 A_{312}; E_{61} = \lambda^4 A_{612}; E_{91} = \lambda^4 A_{912}; E_{121} = \lambda^4 A_{1212}$$
Eq. (A.11)

**Matrix F**

$$F=\begin{pmatrix} 1-F_{11} & 0 \\ -1-F_{11} & 0 \\ 0 & \frac{12}{e^1}(1+F_{32}) \\ F_{41} & 0 \\ -F_{51} & 0 \\ 0 & -\frac{12}{e^2}F_{62} \\ -F_{71} & 0 \\ F_{81} & 0 \\ 0 & -\frac{12}{e^3}F_{92} \\ -F_{101} & 0 \\ F_{101} & 0 \\ 0 & \frac{12}{e^4}F_{122} \end{pmatrix} \quad G=\begin{pmatrix} G_{11} & 0 \\ G_{11} & 0 \\ 0 & -\frac{12}{e^1}G_{32} \\ G_{41} & 0 \\ G_{51} & 0 \\ 0 & \frac{12}{e^2}G_{62} \\ -G_{71} & 0 \\ -G_{81} & 0 \\ 0 & -\frac{12}{e^3}F_{92} \\ -1+G_{101} & 0 \\ -1-G_{101} & 0 \\ 0 & -\frac{12}{e^4}(1+G_{122}) \end{pmatrix} \qquad \text{Eq. (A.12)}$$

The 10 $F_{jk}$ and 10 $G_{jk}$ coefficients are:

$$F_{11} = \lambda^1 a_{33}; \; F_{32} = \left(\frac{9}{2}\frac{\kappa^1}{35}\right)C_{33}; \; F_{41} = \frac{\lambda^1}{12}B_{34}; \; F_{51} = \lambda^1 A_{36}; \; F_{62} = \left(\frac{9}{2}\frac{\kappa^1}{35}\right)C_{36}$$
$$F_{71} = \frac{\lambda^1}{12}B_{37}; \; F_{81} = \lambda^1 A_{39}; \; F_{92} = \left(\frac{9}{2}\frac{\kappa^1}{35}\right)C_{39}; \; F_{101} = \lambda^1 A_{312}; \; F_{122} = \left(\frac{9}{2}\frac{\kappa^1}{35}\right)C_{312}$$
$$G_{11} = \lambda^4 A_{312}; \; G_{32} = \left(\frac{9}{2}\frac{\kappa^4}{35}\right)C_{312}; \; G_{41} = \frac{\lambda^4}{12}B_{412}; \; G_{51} = \lambda^4 A_{612}; \; G_{62} = \left(\frac{9}{2}\frac{\kappa^4}{35}\right)C_{612}$$
$$G_{71} = \frac{\lambda^4}{12}B_{712}; \; G_{81} = \lambda^4 A_{912}; \; G_{92} = \left(\frac{9}{2}\frac{\kappa^4}{35}\right)C_{912}; \; G_{101} = \lambda^4 a_{1212}; \; G_{122} = \left(\frac{9}{2}\frac{\kappa^4}{35}\right)C_{1212}$$
Eq. (A.13)

# Appendix B - Boundary conditions

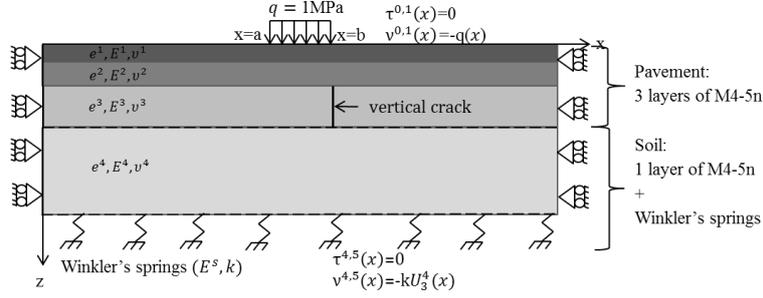

**At the edges**

$$Cx_1 X'(x) + Cx_2 X(x) = Cx_3 Y^{0,1}(x) + Cx_4 Y^{4,5}(x) \quad \text{Eq. (B.1)}$$

$$Cx_1 = \begin{pmatrix} 0 & 0 & 0 & 0 & 0 & 0 & 0 & 0 & 0 & 0 & 0 & 0 \\ 0 & 0 & 0 & 0 & 0 & 0 & 0 & 0 & 0 & 0 & 0 & 0 \\ 0 & 0 & e^1 A_{33} & 0 & 0 & e^2 A_{36} & 0 & 0 & -e^3 A_{39} & 0 & 0 & -e^4 A_{312} \\ 0 & 0 & 0 & 0 & 0 & 0 & 0 & 0 & 0 & 0 & 0 & 0 \\ 0 & 0 & 0 & 0 & 0 & 0 & 0 & 0 & 0 & 0 & 0 & 0 \\ 0 & 0 & e^1 A_{36} & 0 & 0 & e^2 A_{66} & 0 & 0 & -e^3 A_{69} & 0 & 0 & -e^4 A_{612} \\ 0 & 0 & 0 & 0 & 0 & 0 & 0 & 0 & 0 & 0 & 0 & 0 \\ 0 & 0 & 0 & 0 & 0 & 0 & 0 & 0 & 0 & 0 & 0 & 0 \\ 0 & 0 & -e^1 A_{39} & 0 & 0 & -e^2 A_{69} & 0 & 0 & e^3 A_{99} & 0 & 0 & e^4 A_{912} \\ 0 & 0 & 0 & 0 & 0 & 0 & 0 & 0 & 0 & 0 & 0 & 0 \\ 0 & 0 & 0 & 0 & 0 & 0 & 0 & 0 & 0 & 0 & 0 & 0 \\ 0 & 0 & -e^1 A_{312} & 0 & 0 & -e^2 A_{612} & 0 & 0 & e^3 A_{912} & 0 & 0 & e^4 A_{1212} \end{pmatrix} \quad \text{Eq. (B.2)}$$

$$Cx_2 = \begin{pmatrix} 1 & 0 & 0 & 0 & 0 & 0 & 0 & 0 & 0 & 0 & 0 & 0 \\ 0 & 1 & 0 & 0 & 0 & 0 & 0 & 0 & 0 & 0 & 0 & 0 \\ -B_{13} & -\frac{5e^1}{12}B_{23} & 0 & B_{34} & -\frac{5e^2}{12}B_{16} & 0 & -B_{37} & \frac{5e^3}{12}B_{19} & 0 & -B_{112} & \frac{5e^4}{12}B_{112} & 0 \\ 0 & 0 & 0 & 1 & 0 & 0 & 0 & 0 & 0 & 0 & 0 & 0 \\ 0 & 0 & 0 & 0 & 1 & 0 & 0 & 0 & 0 & 0 & 0 & 0 \\ -B_{16} & -\frac{5e^1}{12}B_{16} & 0 & B_{46} & -\frac{5e^2}{12}B_{56} & 0 & -B_{67} & \frac{5e^3}{12}B_{59} & 0 & -B_{512} & \frac{5e^4}{12}B_{512} & 0 \\ 0 & 0 & 0 & 0 & 0 & 1 & 0 & 0 & 0 & 0 & 0 & 0 \\ 0 & 0 & 0 & 0 & 0 & 0 & 1 & 0 & 0 & 0 & 0 & 0 \\ B_{19} & \frac{5e^1}{12}B_{19} & 0 & B_{49} & \frac{5e^2}{12}B_{59} & 0 & -B_{79} & -\frac{5e^3}{12}B_{89} & 0 & B_{812} & \frac{-5e^4}{12}B_{812} & 0 \\ 0 & 0 & 0 & 0 & 0 & 0 & 0 & 0 & 1 & 0 & 0 & 0 \\ 0 & 0 & 0 & 0 & 0 & 0 & 0 & 0 & 0 & 1 & 0 & 0 \\ B_{112} & \frac{5e^1}{12}B_{112} & 0 & B_{412} & \frac{5e^2}{12}B_{512} & 0 & -B_{712} & \frac{-5e^3}{12}B_{812} & 0 & B_{1012} & \frac{-5e^4}{12}B_{1112} & 0 \end{pmatrix} \quad \text{Eq. (B.3)}$$

$$Cx_3 = \begin{pmatrix} 0 & 0 \\ 0 & 0 \\ -1 - D_{31} & 0 \\ 0 & 0 \\ 0 & 0 \\ -D_{61} & 0 \\ 0 & 0 \\ 0 & 0 \\ D_{91} & 0 \\ 0 & 0 \\ 0 & 0 \\ D_{121} & 0 \end{pmatrix} \quad Cx_4 = \begin{pmatrix} 0 & 0 \\ 0 & 0 \\ E_{31} & 0 \\ 0 & 0 \\ 0 & 0 \\ E_{61} & 0 \\ 0 & 0 \\ 0 & 0 \\ -E_{91} & 0 \\ 0 & 0 \\ 0 & 0 \\ -1 - E_{121} & 0 \end{pmatrix} \quad \text{Eq. (B.4)}$$

## Vertical crack across the third layer

$$Cx_{1fiss}X'(x) + Cx_{2fiss}X(x) = Cx_{3fiss}Y^{0,1}(x) + Cx_{4fiss}Y^{4,5}(x) \qquad \text{Eq. (B.5)}$$

$$Cx_{1fiss} = \begin{pmatrix}
0 & 0 & 0 & 0 & 0 & 0 & 0 & 0 & 0 & 0 & 0 & 0 \\
0 & 0 & 0 & 0 & 0 & 0 & 0 & 0 & 0 & 0 & 0 & 0 \\
0 & 0 & e^1 A_{33} & 0 & 0 & e^2 A_{36} & 0 & 0 & -e^3 A_{39} & 0 & 0 & -e^4 A_{312} \\
0 & 0 & 0 & 0 & 0 & 0 & 0 & 0 & 0 & 0 & 0 & 0 \\
0 & 0 & 0 & 0 & 0 & 0 & 0 & 0 & 0 & 0 & 0 & 0 \\
0 & 0 & e^1 A_{36} & 0 & 0 & e^2 A_{66} & 0 & 0 & -e^3 A_{69} & 0 & 0 & -e^4 A_{612} \\
0 & 0 & 0 & 0 & 0 & 0 & 1 & 0 & 0 & 0 & 0 & 0 \\
0 & 0 & 0 & 0 & 0 & 0 & 0 & 1 & 0 & 0 & 0 & 0 \\
0 & 0 & -e^1 A_{39} & 0 & 0 & -e^2 A_{69} & 0 & 0 & e^3 A_{99} & 0 & 0 & e^4 A_{912} \\
0 & 0 & 0 & 0 & 0 & 0 & 0 & 0 & 0 & 0 & 0 & 0 \\
0 & 0 & 0 & 0 & 0 & 0 & 0 & 0 & 0 & 0 & 0 & 0 \\
0 & 0 & -e^1 A_{312} & 0 & 0 & -e^2 A_{612} & 0 & 0 & e^3 A_{912} & 0 & 0 & e^4 A_{1212}
\end{pmatrix} \qquad \text{Eq. (B.6)}$$

$$Cx_{2fiss} = \begin{pmatrix}
1 & 0 & 0 & 0 & 0 & 0 & 0 & 0 & 0 & 0 & 0 & 0 \\
0 & 1 & 0 & 0 & 0 & 0 & 0 & 0 & 0 & 0 & 0 & 0 \\
-B_{13} & -\frac{5e^1}{12}B_{23} & 0 & B_{34} & -\frac{5e^2}{12}B_{16} & 0 & -B_{37} & \frac{5e^3}{12}B_{19} & 0 & -B_{112} & \frac{5e^4}{12}B_{112} & 0 \\
0 & 0 & 0 & 1 & 0 & 0 & 0 & 0 & 0 & 0 & 0 & 0 \\
0 & 0 & 0 & 0 & 1 & 0 & 0 & 0 & 0 & 0 & 0 & 0 \\
-B_{16} & -\frac{5e^1}{12}B_{16} & 0 & B_{46} & -\frac{5e^2}{12}B_{56} & 0 & -B_{67} & \frac{5e^3}{12}B_{59} & 0 & -B_{512} & \frac{5e^4}{12}B_{512} & 0 \\
0 & 0 & 0 & 0 & 0 & 0 & 0 & 0 & 0 & 0 & 0 & 0 \\
0 & 0 & 0 & 0 & 0 & 0 & 0 & 0 & 0 & 0 & 0 & 0 \\
B_{19} & \frac{5e^1}{12}B_{19} & 0 & B_{49} & \frac{5e^2}{12}B_{59} & 0 & -B_{79} & -\frac{5e^3}{12}B_{89} & 0 & B_{812} & \frac{-5e^4}{12}B_{812} & 0 \\
0 & 0 & 0 & 0 & 0 & 0 & 0 & 0 & 0 & 1 & 0 & 0 \\
0 & 0 & 0 & 0 & 0 & 0 & 0 & 0 & 0 & 0 & 1 & 0 \\
B_{112} & \frac{5e^1}{12}B_{112} & 0 & B_{412} & \frac{5e^2}{12}B_{512} & 0 & -B_{712} & \frac{-5e^3}{12}B_{812} & 0 & B_{1012} & \frac{-5e^4}{12}B_{1112} & 0
\end{pmatrix} \qquad \text{Eq. (B.7)}$$

$$Cx_{3fiss} = Cx_3 \; ; \; Cx_{4fiss} = Cx_4 \qquad \text{Eq. (B.8)}$$